\documentclass[11pt]{article}
\usepackage{geometry}                
\usepackage{graphicx}
\usepackage{amssymb}
\usepackage{cite}
\usepackage{epstopdf}
\usepackage{mathtools}
\usepackage{fullpage}
\usepackage{subfig}
\usepackage{bm}
\usepackage{lscape}
\usepackage{booktabs}
\usepackage{pdflscape}
\usepackage[english]{babel}
\usepackage{adjustbox}
\usepackage{datetime}
\usepackage{authblk}

\usepackage[compact]{titlesec}

\AtBeginDocument{}

\title{\huge Complex Politics: A Quantitative Semantic and Topological Analysis of UK House of Commons Debates\thanks{This paper has been designed during the 2015 Santa Fe Institute Complex Systems Summer School. Data pre-processing and post-processing had been performed by Stefano Gurciullo and Michael Smallegan. Dynamic Topic Modeling has been done by Sebastian Poledna, while Alice Patania dealt with Topological Data Analysis. Daniel Hedblom interpreted DTM results, Federico Battiston processed them to obtain insights for individual political agents. Model selection was done by Marìa Pereda Garcia and Bahattin Tolga Oztan. The manuscript has been written by Stefano Gurciullo, Michael Smallegan, Marìa Pereda Garcia, Daniel Hedblom, Federico Battiston, Bahattin Tolga Oztan. Alexander Herzog, Peter John, Slava Mikhaylov are respondible for the data extraction and the preliminary design of the paper. Coordination and team tasks have been managed by Stefano Gurciullo. We thank Nilton Cardoso for his support during the preliminary implementation of the research project. Our gratitude goes also towards the organizers, faculty and participants of the Summer School, who provided an inspirational context and valuable feedback for our project. Correspondence regarding the paper can be sent to stefano.gurciullo.11@ucl.ac.uk.}}

\author[1]{Stefano Gurciullo}
\author[2]{Michael Smallegan}
\author[3]{Mar\'{i}a Pereda}
\author[4]{Federico Battiston}
\author[5]{Alice Patania}
\author[6]{Sebastian Poledna}
\author[7]{Daniel Hedblom}
\author[8]{Bahattin Tolga Oztan}
\author[9]{Alexander Herzog}
\author[1]{Peter John}
\author[1]{Slava Mikhaylov}

\affil[1]{School of Public Policy, University College London}
\affil[2]{Wisconsin Institute for Discovery, University of Wisconsin-Madison}
\affil[3]{University of Burgos}
\affil[4]{Queen Mary, University of London}
\affil[5]{Politecnico di Torino, I{.}S{.}I Foundation}
\affil[6]{Medical University of Vienna}
\affil[7]{The University of Chicago}
\affil[8]{UCI School of Social Sciences}
\affil[9]{Social Analytics Institute and Big Data Systems Lab, Clemson University}

\newdate{date}{05}{10}{2015}
\date{\displaydate{date}}

\begin{document}

\maketitle

\vspace{-12mm}

\begin{abstract}
    \centering
    \begin{minipage}{0.9\textwidth}
\linespread{1.2}\small \textbf{Abstract:} This study is a first, exploratory attempt to use quantitative semantics techniques and topological analysis to analyze systemic patterns arising in a complex political system. In particular, we use a rich data set covering all speeches and debates in the UK House of Commons between 1975 and 2014. By the use of dynamic topic modeling (DTM) and topological data analysis (TDA) we show that both members and parties feature specific roles within the system, consistent over time, and extract global patterns indicating levels of political cohesion. Our results provide a wide array of novel hypotheses about the complex dynamics of political systems, with valuable policy applications.
    \end{minipage}
\end{abstract}

\vspace{2mm}

\newpage
\section{Introduction}

Complexity science has grown to become an important paradigm within social science. Seminal work, ranging from Axelrod's studies of the emergence of cooperation \cite{axelrod1981evolution} to Arthur's conceptualization of the economy as a complex system \cite{arthur1999complexity}, has had a significant impact in sociology and economics. Applications of complex systems methodologies in these two fields have made it possible to analyze interesting real world phenomena such as status-seeking in online communities \cite{lampel2007role} and financial shock contagion and impact \cite{battiston2015leveraging} in a way which would have been impossible with traditional tools.

Compared to such advances, political science and public policy studies appear to have been less exposed to complexity science. While theoretical studies depicting political systems as complex system do exist \cite{cairney2012complexity}, empirical investigations are still few and far between. The reason is straightforward: lack of data has prevented researchers from uncovering the complex dynamics behind political affairs. The trend is, however, changing thanks to novel data sets and advances in the processing of unstructured information.

This study is a first, exploratory attempt to contribute to this line of work. We investigate systemic patterns arising in the UK House of Commons by applying quantitative semantics techniques and topological analysis on its debates between 1975 and 2014. We discover that both members and parties feature specific roles within the system, consistent over time and providing room for novel hypotheses about the complex dynamics of political systems.

Four more sections follow these paragraphs. First, we contextualize our research on the grounds of past scholarly attempts of defining political institutions and policy as complex systems. Second, we introduce the Hansard political speeches dataset, presenting its key summary statistics and features. The section is followed by an explanation of the methods used to extract patterns from the data, namely Dynamic Topic Modelling (DTM) and Topological Data Analysis (TDA). Section 5 shows and interprets the results of this work, with an elaboration of potential hypotheses explaining them. We conclude by summarizing our findings and providing routes for future work. 

%
%

\section{Background: Complexity and Political Science}

%
%


The \textit{structuralist} perspective is a well established paradigm in political theory. It views politics as a system of interacting agents aimed at the distribution of power and the management of some aspects of public social life \cite{easton1965framework, hay1998structure}. Since its beginning, the structuralist school has experienced several declinations \cite{young1981post, sunkel1990neo}, yet it has not met formally complex systems studies until recently. Jervis \cite{jervis1998system} and Rhee \cite{rhee2000complex} are among the first to assume that political life behaves as a self-organizing system pressured by macro-evolutionary forces. According to these scholars, political institutions change both their nature and surroundings to strive, influenced by the aggregate dynamics and interests of the actors who form them and the stakeholders of their decision-making process.

The view is indeed interesting and worth exploring, yet current endeavors have only fueled a narrative for general recommendations for the policy maker; little progress has been made in terms of quantitative analysis and rigorous testing of hypotheses. The main reason for this is that it hard to empirically identify policy effects in the presence of non-linear feedback phenomena \cite{teisman2008complexity}. Instead, the practitioner is advised not to rely on a single strategy but to diversify their policy actions \cite{geyer2010complexity, cairney2011understanding}.

Of more interest to this paper are the attempts to account for policy dynamics with the punctuated equilibrium hypothesis. Works by Baumgartner and Jones \cite{baumgartner2010agendas} and Workman et al. \cite{workman2009information} suggest that policy actors are surrounded by an enormous amount of signals that is relevant for their decision-making processes. Yet, such actors are affected by cognitive constraints that render them boundedly rational \cite{simon2000bounded}, leading them to perform their decision-making by ignoring most of the signals and concentrate only on a few. The result is a `policy punctuated' type of dynamics: the attention of policy makers would be strictly focused on a few issues, with minimal attention to others \cite{cairney2012complexity}. Sudden changes followed by systemic positive feedback (the so-called 'bandwagon effects' \cite{henshel1987emergence, schmitt2008bandwagon} would cause shifts of priorities in policy agendas. The cause of such mutations may originate from any of the external signals that actors are subject to: economic trends, opinion polls, new governmental appointments. 

The frameworks mentioned are valuable in understanding policy dynamics from a complex systems narrative, however, few are the efforts to empirically validate them, being limited mainly to qualitative investigations \cite{geyer2012can}. Differently from other social sciences, the study of public policy eschewed the possibility to gather the right kind of data to quantitatively understand these phenomena. 

In this paper we propose an approach to explore policy dynamics - thus allowing the testing of hypotheses on its nature and patterns - based on the processing of information coming from unstructured textual data. We focus on a specific political institution, the House of Commons of the United Kingdom.

On a first inspection, the House of Commons may appear as a very structured, hierarchical institution. Top-down organization in the form of agenda setting and debate management would make its legislative power quite orderly. In line with the literature cited, we challenge this view by stating that an orderly hierarchical structure coexists with a complex dynamics of political interests, that eventually are crystallized in the form of policy actions through bills. Each member of the House possesses a unique set of interests and preferred political issues, which are dispersed through information dissemination and assimilation. We identify and measure those by analyzing their very speeches and debates, contained in a novel data set presented in the next section.

\section{The data}

Our data is composed of the minutes of all debates occurred at the UK House of Commons between the years 1974-2014. These are extracted from a larger dataset that features speeches dating back from 1935. As outlined in Escher \cite{escher2011theyworkforyou}, all the information has been extracted from a transparency website known as TheyWorkForYou.com, which provides access to UK parliamentary records and other Member of Parliament (MP) information. The information has been downloaded in XML format and processed in Python, where all parts of speech but names and adjectives have been removed. Overall, the dataset contains over 3.7 million individual speeches. In order to allow a dynamic modeling of the data, we divide the observations by parliamentary sessions, which commence with the initial speech of the Queen to the House and ranging about 12 calendar months. We obtain a total of 37 sessions, with an average of about 4800 contributions per session. The number of parliamentary speakers revolves around 630 at each time unit.

To have a sense of the nature of the speeches, Table 1 shows their median length (measured in number of words), as well as the $10^{th}$ and $90^{th}$ percentiles. It can be observed that their length distributions start being skewed more to the right during the 2001-2002 session, accompanied by an increase in the median. The novel pattern indicates a phase shift in the parliamentary activities, perhaps first caused by the concern for post-9/11 terrorism activities, then to economic and social issues affecting the country during the periods pre- and post-financial and debt crisis.

We assume that parliamentary speeches are manifestations of the political interests of the speakers. This implies that by extracting and analyzing their semantic information - i.e. knowing what and how much a speaker talks about a determined policy topic in relation to others - we are able to attain insights on their roles and attitudes within the House of Commons decision-making system. Our goal is reached with the application of dynamic topic modeling and topological data analysis on the dataset, as illustrated in the following section.

\begin{landscape}

\begin{table}[htbp]
  \tiny
    \begin{tabular}{rrrrrrr}
    \toprule
    \multicolumn{1}{c}{\textbf{Session ID}} & \multicolumn{1}{c}{\textbf{Year}} & \multicolumn{1}{c}{\textbf{Number of Speakers}} & \multicolumn{1}{c}{\textbf{Number of speeches}} & \multicolumn{1}{c}{\textbf{Median length of speeches }} & \multicolumn{1}{c}{\textbf{10th percentile}} & \multicolumn{1}{c}{\textbf{90th percentile}} \\
    \midrule
    4701  & 1974-1975 & 623   & 57299 & 152   & 24    & 330 \\
    4702  & 1975-1976 & 627   & 59699 & 149   & 22    & 321 \\
    4703  & 1976-1977 & 618   & 49036 & 155   & 23    & 326 \\
    4704  & 1977-1978 & 626   & 55888 & 154   & 22    & 323 \\
    4705  & 1978-1979 & 583   & 24338 & 162   & 25    & 331 \\
    4802  & 1980-1981 & 617   & 46559 & 157   & 21    & 331 \\
    4803  & 1981-1982 & 624   & 51475 & 154   & 22    & 331 \\
    4804  & 1982-1983 & 608   & 34148 & 153   & 20    & 332 \\
    4901  & 1983-1984 & 644   & 67329 & 152   & 20    & 335 \\
    4902  & 1984-1985 & 640   & 57241 & 155   & 19    & 342 \\
    4903  & 1985-1986 & 639   & 57578 & 158   & 20    & 345 \\
    4904  & 1986-1987 & 628   & 34633 & 153   & 17    & 349 \\
    5001  & 1987-1988 & 642   & 71333 & 157   & 17    & 349 \\
    5002  & 1988-1989 & 647   & 57016 & 160   & 17    & 360 \\
    5003  & 1989-1990 & 648   & 53349 & 163   & 17    & 358 \\
    5004  & 1990-1991 & 642   & 50259 & 166   & 19    & 361 \\
    5005  & 1991-1992 & 614   & 22313 & 157   & 17    & 355 \\
    5101  & 1992-1993 & 645   & 70305 & 169   & 19    & 356 \\
    5102  & 1993-1994 & 644   & 45979 & 160   & 17    & 338 \\
    5103  & 1994-1995 & 642   & 45202 & 165   & 18    & 345 \\
    5104  & 1995-1996 & 629   & 43544 & 168   & 18    & 353 \\
    5105  & 1996-1997 & 609   & 24329 & 169   & 18    & 358 \\
    5201  & 1997-1998 & 652   & 74796 & 169   & 18    & 358 \\
    5202  & 1998-1999 & 644   & 47767 & 172   & 17    & 369 \\
    5203  & 1999-2000 & 644   & 54965 & 166   & 17    & 362 \\
    5204  & 2000-2001 & 621   & 25692 & 176   & 17    & 379 \\
    5301  & 2001-2002 & 645   & 55227 & 273   & 28    & 1572 \\
    5302  & 2002-2003 & 645   & 52591 & 256   & 30    & 1330 \\
    5303  & 2003-2004 & 636   & 47947 & 259   & 34    & 1400 \\
    5304  & 2004-2005 & 611   & 20549 & 262   & 31    & 1379.4 \\
    5401  & 2005-2006 & 634   & 64507 & 254   & 35    & 1269 \\
    5402  & 2006-2007 & 628   & 40142 & 264   & 39    & 1356.9 \\
    5403  & 2007-2008 & 630   & 50175 & 254   & 30    & 1234.6 \\
    5404  & 2008-2009 & 622   & 39883 & 253   & 35    & 1322 \\
    5405  & 2009-2010 & 596   & 21452 & 239   & 36    & 1230 \\
    5502  & 2012-2013 & 639   & 53975 & 222   & 44    & 1104.6 \\
    5503  & 2013-2014 & 634   & 46467 & 225   & 46    & 1165 \\
          &       &       &       &       &       &  \\
          \toprule
          & \textbf{Average} & 630   & 47973 & 187   & 24    & 631 \\
    \bottomrule
    \end{tabular}%
  \label{tab:addlabel}%
  \centering
  \caption{UK House of Commons Debates - Summary statistics.}
\end{table}%

\end{landscape}

\section{Methods}
\subsection{Dynamic Topic Modeling}
The political interests of House of Commons speakers are latent variables which, we assume, manifest themselves in the debates. Traditional approaches in the social sciences would attempt to unearth them through methods such as word counts and tag clouds \cite{collins2009parallel, ramage2009topic}. These would not yield valuable insights for our goal, as they are not able to provide a clear-cut, definite range of semantic sets - policy topics - that change over time. For this reason, we adopt an unsupervised machine learning approach known as Dynamic Topic Modeling (DTM). 

Topic models are a family of generative probabilistic models aimed at classifying co-occurring words in text corpora into specific groups or distributions \cite{wallach2006topic, blei2012probabilistic}. DTM, more specifically, captures the evolution of topics in a sequentially organized corpus of documents. Given a $T$ number of topics pre-defined by the user, DTM assigns the probability that a words appears in any of them by the use of a state space model, which incorporates assumptions about the shape of the topics distributions. For a detailed mathematical account of the model and the respective algorithm, see \cite{blei2006dynamic}. \\

Empirical applications of topic models have shown that the inferred topic distributions often feature semantically valuable content \cite{hong2010empirical}. In a previous work, the application of Latent Dirichlet Allocation, a specific kind of topic model, on House of Commons speeches has demonstrated that the resulting topics conform very closely with policy categories as produced by human-based content analysis \cite{gurciullo2014policy}. We depart from this finding and extend it to identify policy topics and how their composition and nature mutate over time. The work flow is as follows: 

\begin{enumerate}
  \item The corpus of speeches is split into 37 time slices, corresponding to the parliamentary sessions. For each session, all speeches made by the same MP are aggregated into a single document. 
  \item A vocabulary and a term-document matrix are produced out of the corpus, and fed to the DTM script. The number of topics $T$ is 15. Appendix A explains the cross-validation approach followed to determine it.
  \item DTM results are used to evaluate a dynamic probability vector for each MP at each time slice. In other words, for each session that a speaker takes part to, the probability that his speeches are classified to any of the 15 topics is calculated. 
  \item House of Commons speakers' probability vectors are analyzed to identify patterns and roles across individuals and parties.
\end{enumerate}

The corpus pre- and post-processing is performed in Python. The DTM model is run on a C script.

\subsection{Topological Data Analysis}
Topological Data Analysis (TDA) is a relatively new area of research first introduced in 2002 by Gunnar Carlsson \cite{carlsson_ams}. The basic assumption of TDA is that any kind of data can be seen as a sampling of a manifold, which can be studied using topological tools that are sensitive to both large and small scale patterns. 
In this study we decided to use a partial clustering method based on the Mapper algorithm first introduced in \cite{singh}
and later commercialized by the company Ayasdi whose main product has the Mapper algorithm at its core. This algorithm has been used before to study voting behaviors of the U.S. House of Representatives in \cite{mapper}.

The basic idea of the Mapper algorithm is to perform clustering across different scales, and then track how these clusters change as the scale varies. To do so a distance and a filtering function are defined on the data. The procedure of Mapper is very simple. At first, an open covering of the data set $X$ is constructed according to the filtering function $f:X\to \mathrm{R}$. The image $f(X)$ is divided into intervals $I_k$ of the same range $\rho$, such that each interval overlaps with the consequent one.
Afterwards the distance function is used to cluster the subsets of data in $X_k=f^{-1}(I_k)$, and each cluster is represented in the constructed network by a single point. Edges in the network represent clusters in consequent bins that have points in common.

To capture the temporal dynamics of the speakers,  we used the Mapper algorithm with time as a filter function to construct a network representing the main features of the dataset. The choice of a discrete filter forced us to slightly adjust the algorithm: instead of intervals $I_k=\{t_k\}$ we used single time steps. Notice that given the particular nature of our data, the absence of an overlap in the defined intervals did not imply that the subsets $X_k$ had no points in common, since each politician can be identified in more than one session depending on the term he was elected in. 
The choice of the distance function is motivated by the kind of analysis one wants to perform. In the current study we chose the euclidean distance to identify politicians talking about the same topics for a similar time period.
The last step for the topological data analysis is to define the clustering method. As clustering method we used the Affinity Propagation algorithm introduced in \cite{AP} by Frey and Dueck, since it does not require the number of clusters to be determined or estimated before running the algorithm.\\

The Mapper Algorithm was coded in Python and the sklearn module was used to implement the Affinity Propagation algorithm.

\newpage

\section{Results}
\subsection{Topic Distributions}

We now turn to the results from the DTM analysis and the topic distributions. Table 2 presents the results from the DTM analysis by topic and session (years 1974-75). The top panel contains the first session and the bottom panel contains the last session covered by our data set (years 2013-14). Each column represents one specific topic and lists the 15 top words on for each topic in order of descending importance, i.e. the probability of appearing in the topic distribution. For example, the first topic in the first session, labeled ``International Trade'' contains words such as ``state'', ``trade'', and ``oil''. The topics in each column are the most significant topics debated in the UK parliament during the period we study, and will form an integral component in classifying members of the parliament in the section below. 

A few things are important to point out regarding the construction of the table. First, the set of topics stays fixed over time. The DTM estimation selects the top words for each topic and for each session. (See the Appendix for a discussion on how to optimally choose the number of topics) Second, there is no formal method behind the assignment of labels to topics. Instead, we have simply tried to exercise good judgment and have labeled the topics accordingly. 

Given the selection procedure for topics, the table should mainly be thought of as illustrative of the change in word patterns across topics. Moreover, we need to be careful with not inferring too much from the word distributions alone; the main point of the identified topics will be to classify speakers and construct networks. Keeping this mind, we now proceed to examine some of the changes between the first and the last session. In the ``infrastructures'' topic, it is clear that we see a shift from words related to colonies and airline logistics, to a focus on airports and airlines. For ``regional affairs'' we see a shift from quite general words to an apparent focus on words related to Europe and, in particular, to the European Union. One can also observe that among the top words in the ``entertainment and media'' in the first session, we have ``author'', ``local'', and ``land.'' In the last session, the very same topic contains ``pub,'' ``sport,'' ``dog,'' and ``beer.''

\begin{figure}
\begin{center}
\subfloat{\includegraphics[scale=0.3]{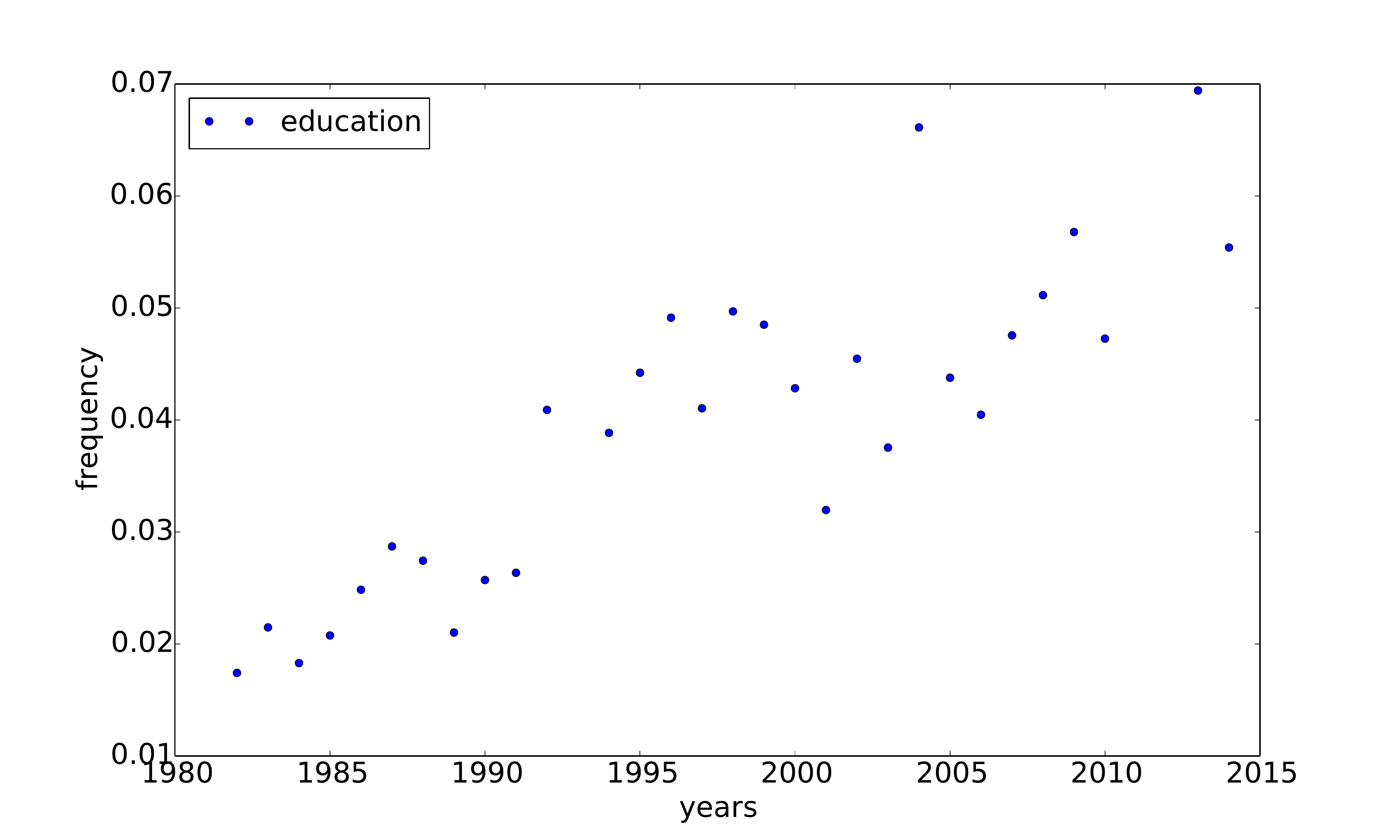}} \\
\subfloat{\includegraphics[scale=0.3]{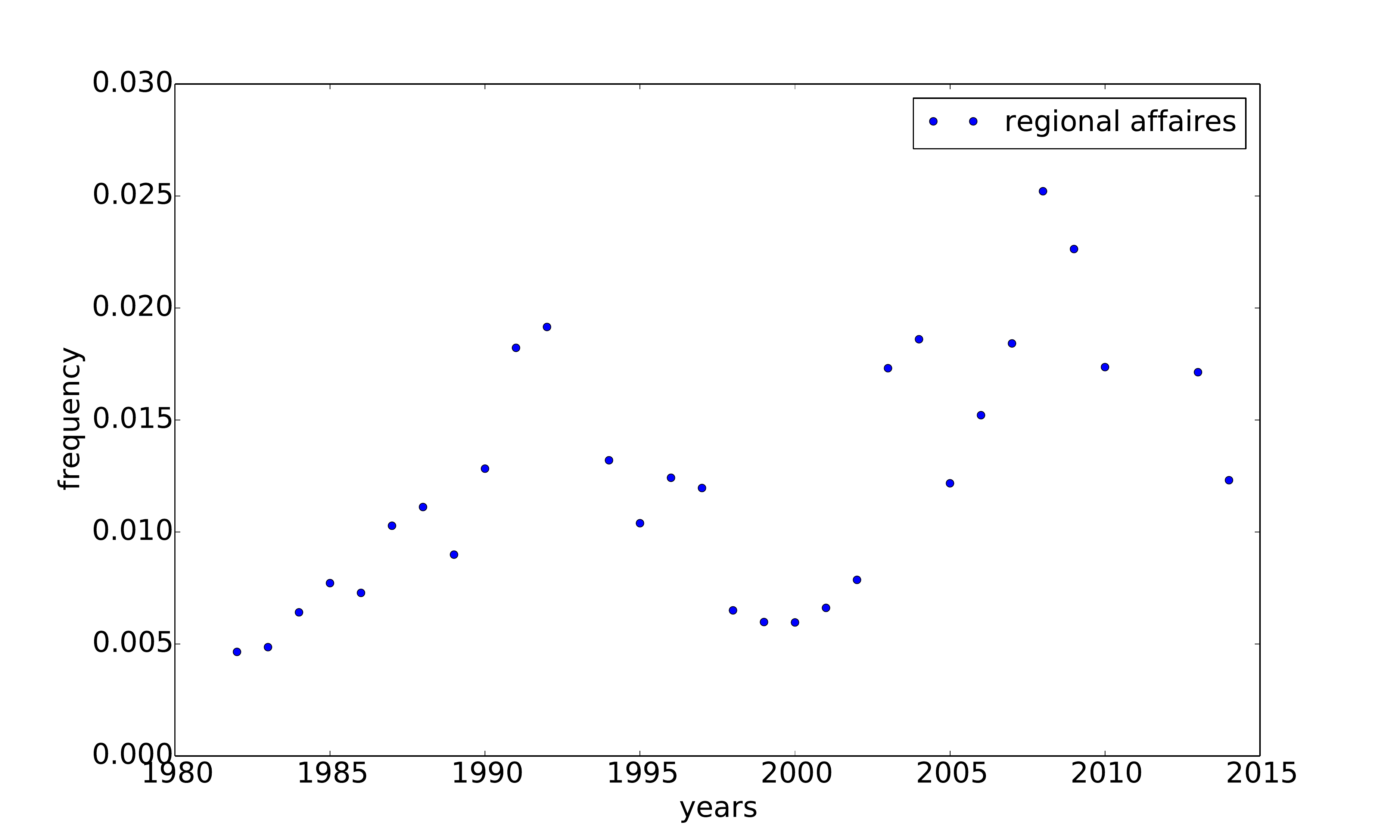}} \\
\subfloat{\includegraphics[scale=0.3]{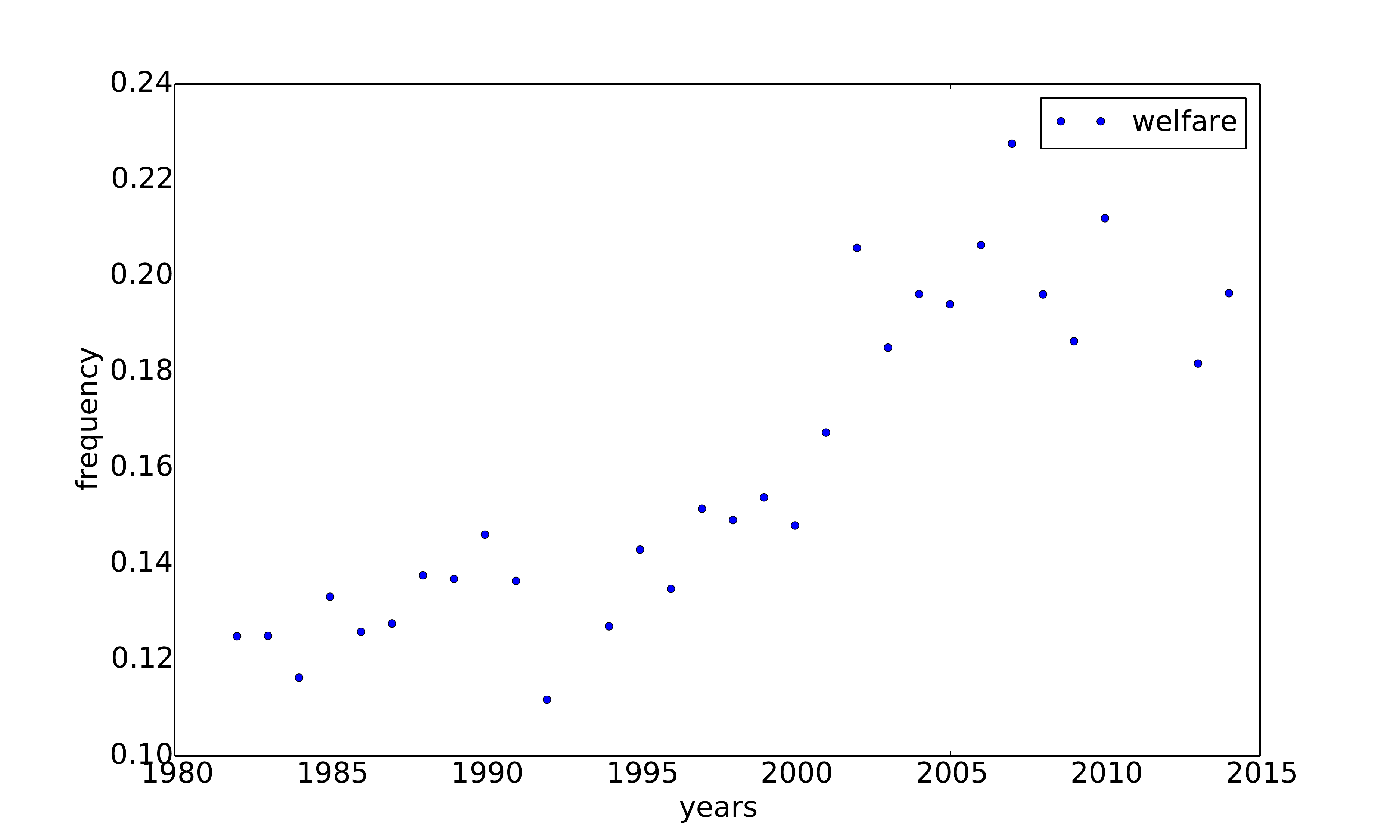}}
\caption[]{Topics with increasing degree of importance over time. Frequency refers to fraction of speeches in which the topic is discussed}
\label{fig:overtime_increase}
\end{center}
\end{figure}

How do the relative importance of different topics evolve over time? Exogenous factors in the economy and society strongly affect which topics speakers spend time on in speeches and debates. At the same time, there is a dynamic interaction of various topics over time and across speakers which causes fluctuations in each topic's relative importance. We will now examine the dynamics of a number of topics which significantly increased or decreased in importance during the period studied. (Note that the remaining topics did not exhibit a clear trend in either direction.) Figure \ref{fig:overtime_increase} shows three topics (from top to bottom): education, regional affairs, and welfare. For education (top figure), we see a fluctuating, but steady increase in importance. The topic ``regional affairs'' (middle panel), on the other hand, increases during the period studied, but exhibits a drop in importance during the late 90s and early 2000s. In contrast, the topic ``welfare'' (bottom figure) increases substantially during the these years, but remained relatively constant during the 80s and 90s.

\begin{figure}
\begin{center}
\subfloat{\includegraphics[scale=0.3]{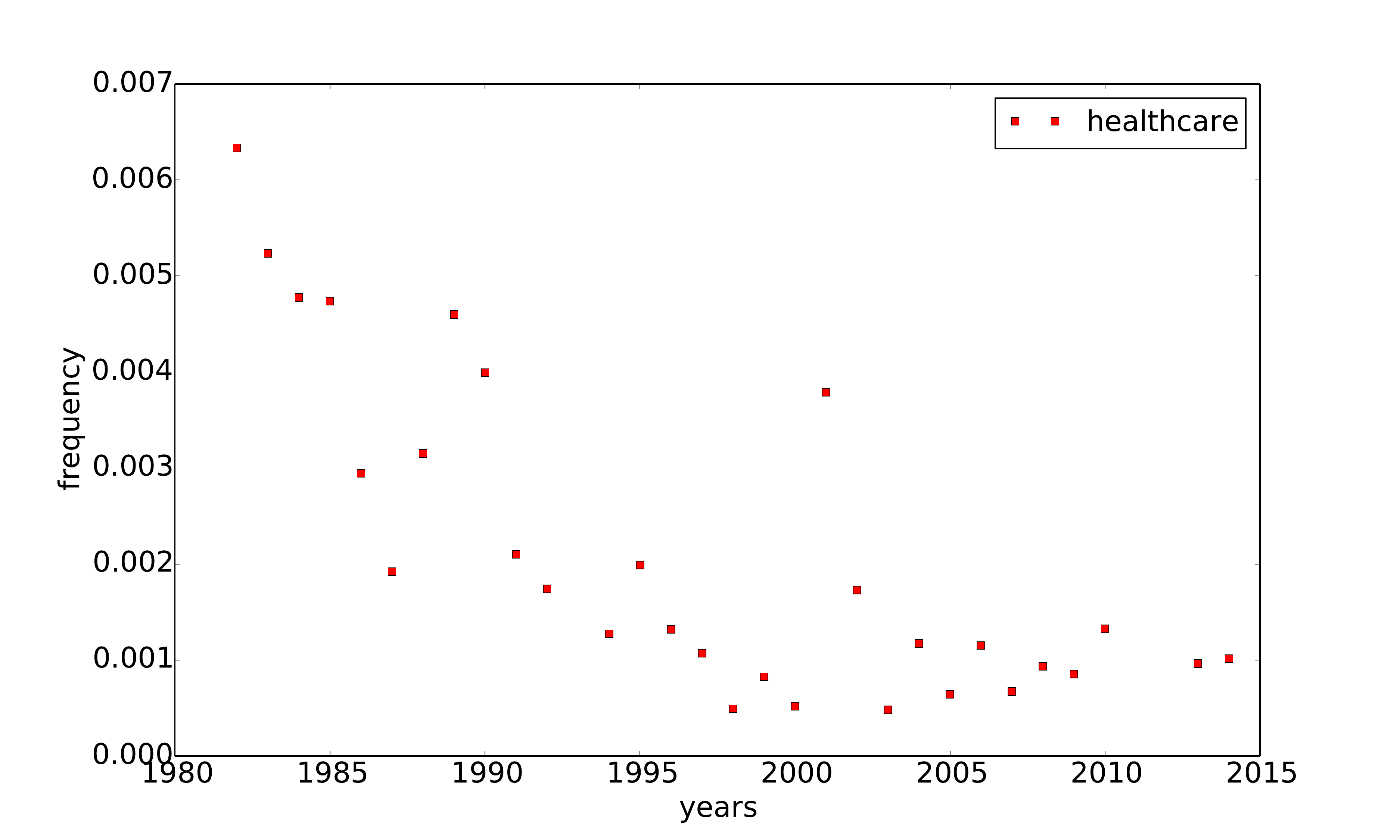}} \\
\subfloat{\includegraphics[scale=0.3]{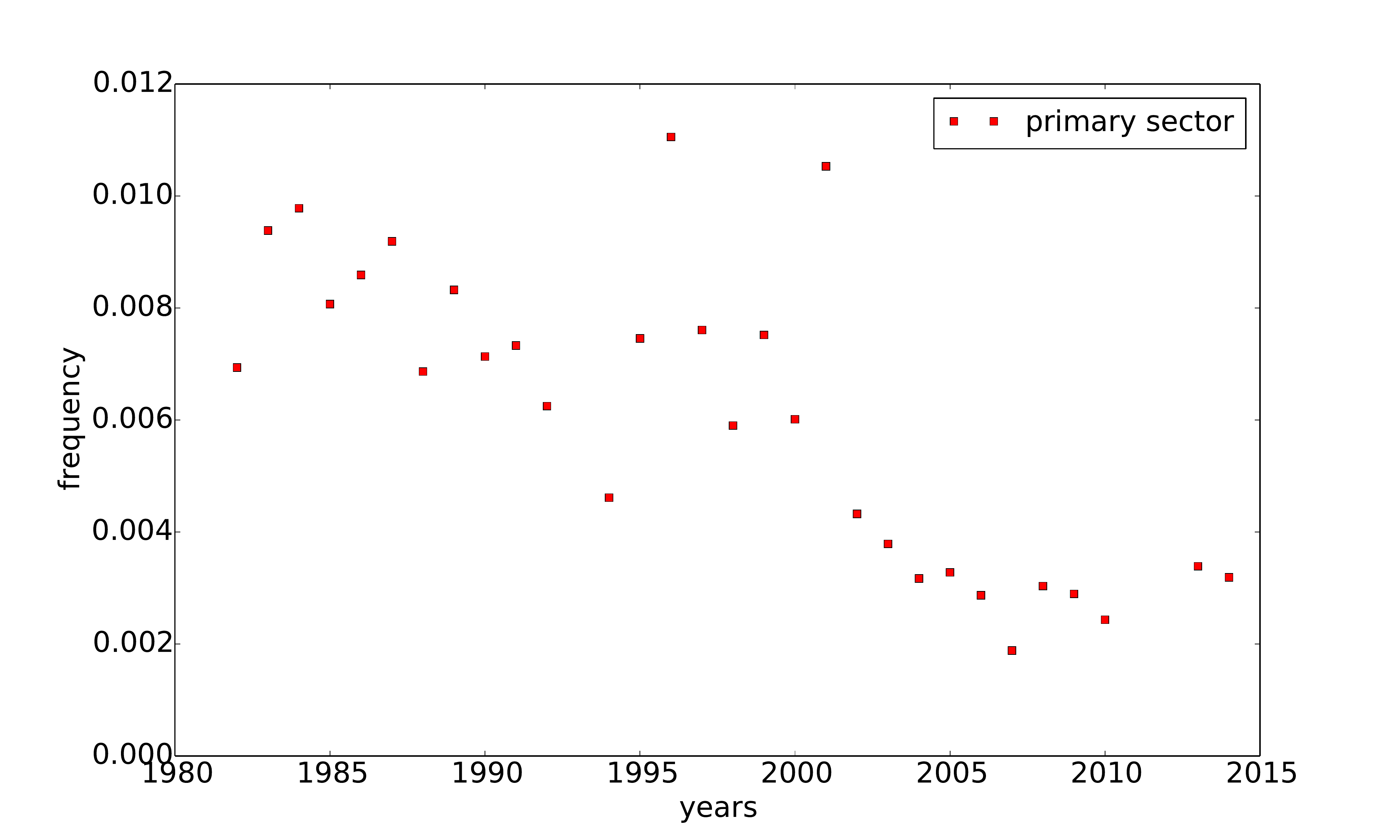}} \\
\subfloat{\includegraphics[scale=0.3]{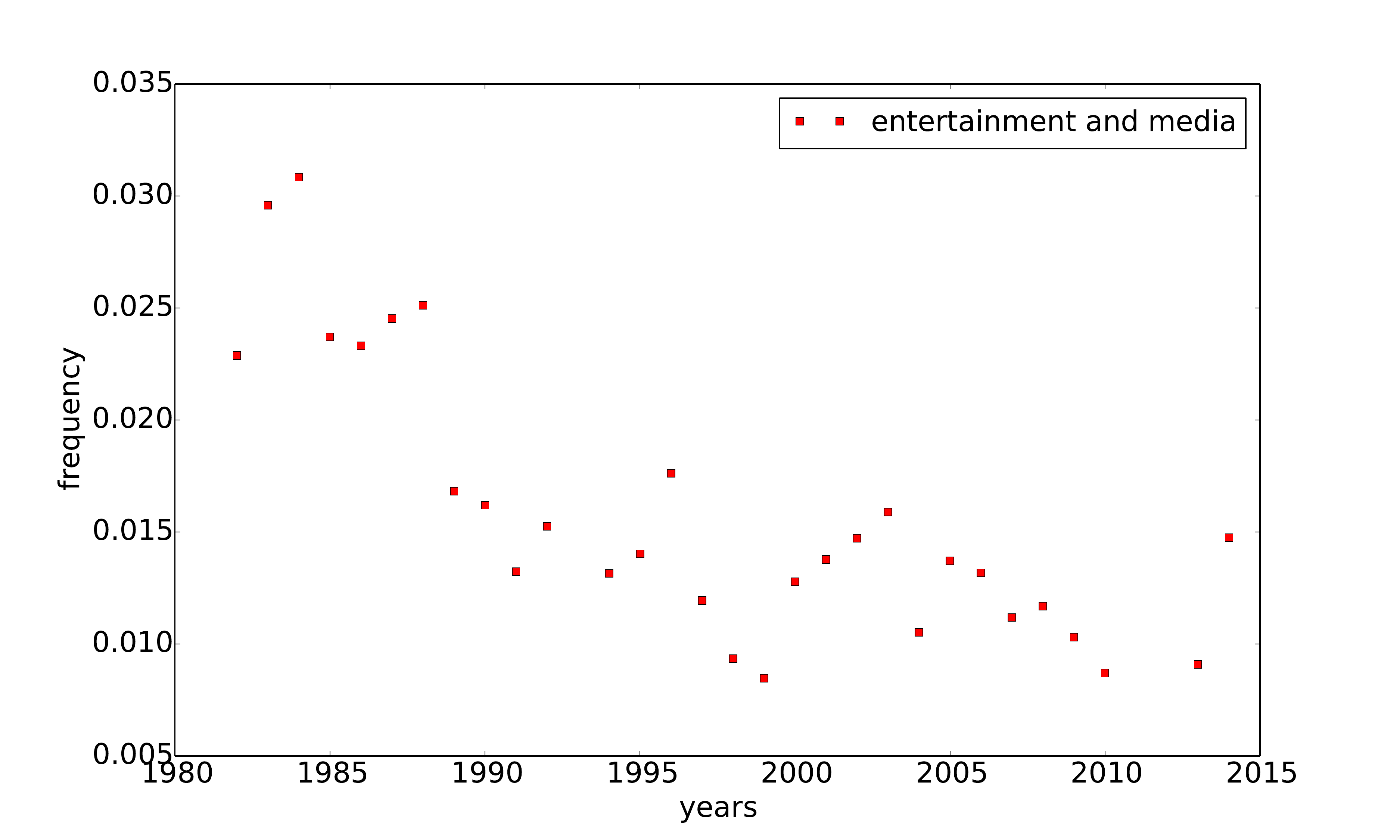}}
\caption[]{Topics with decreasing degree of importance over time. Frequency refers to fraction of speeches in which the topic is discussed}
\label{fig:overtime_decrease}
\end{center}
\end{figure}

Figure \ref{fig:overtime_decrease} shows the three topics with a significant trend of decreasing importance over time. The ``health care'' topic (top figure) decreased substantially in importance during the 80s and early 90s, but remained fairly constant (at a very low level) during the latter part of the period. For the ``primary sector'' topic, we see a constant decline over the whole period. Finally, ``entertainment and media'' exhibits a sudden decline in importance during the late 80s.

\begin{landscape}

\vspace{2mm}

\begin{table}
\tiny
\centering
\begin{adjustbox}{max width=680pt}
\begin{tabular}{ c c c c c c c c c c c c c c c  }

\hline
\hline

International Trade & Education & Healthcare & Procedures & Miscellaneous & Economy & Infrastructures & Welfare & Regional Affairs & Military & Foreign Affairs & Entertainment and Media & Transports & Devolution & Primary sector  \\

\hline

industri & school & health & hous & holder & tax & rhodesia & peopl & area & ireland & countri & author & transport & wale & agricultur \\
state & educ & hospit & point & plot & rate & africa & state & citi & northern & question & local & road & welsh & food \\
price & author & servic & matter & cleethorp & chancellor & aircraft & employ & inner & state & matter & hous & rail & agenc & price \\
secretari & teacher & doctor & amend & aye & inflat & british & secretari & merseysid & secretari & communiti & land & railway & develop & farmer \\
british & local & patient & time & allot & cent & african & servic & liverpool & defenc & hous & council & british & referendum & market \\
compani & children & bed & order & yemen & incom & ship & awar & scotland & forc & foreign & build & fare & state & fisheri \\
polici & parent & medic & committe & yemeni & increas & airport & unemploy & town & peopl & polici & rate & traffic & secretari & fish \\
matter & comprehens & nation & debat & noe & expenditur & port & mani & urban & awar & view & properti & vehicl & devolut & produc \\
trade & colleg & consult & way & plymyard & public & aviat & industri & birmingham & hous & discuss & rent & london & local & milk \\
question & student & profess & question & beg & budget & airlin & area & midland & secur & think & london & servic & water & communiti \\
oil & scienc & nurs & secretari & abil & capit & airway & benefit & region & order & european & water & counti & author & common \\
countri & depart & nhs & peopl & abl & taxat & air & problem & rate & statement & meet & area & bus & counti & pound \\
import & univers & region & state & abolit & money & gime & number & scottish & polic & statement & grant & rural & area & farm \\
hous & system & author & speaker & abrog & excheq & rhodesian & scheme & council & mani & minist & expenditur & car & england & mile \\
time & grant & care & mani & absolut & polici & concord & social & partnership & prison & import & tenant & commut & cardiff & beef \\
\hline
International Trade & Education & Healthcare & Procedures & Miscellaneous & Economy & Infrastructures & Welfare & Regional Affairs &Military & Foreign Affairs & Entertainment and Media & Transports & Devolution & Primary sector \\

\hline
busi & peopl & child & hous & yemen & economi & heathrow & peopl & bank & forc & countri & pub & rail & wale & farmer \\
peopl & work & children & peopl & plot & tax & aviat & local & tax & defenc & european & sport & transport & welsh & anim \\ 
compani & pension & adopt & time & holder & chancellor & air & care & financi & peopl & foreign & industri & line & badger & food \\
energi & tax & famili & point & cleethorp & growth & flight & servic & rate & time & intern & bbc & road & devolut & diseas \\
new & benefit & medicin & committe & allot & budget & runway & health & busi & countri & e & dog & london & rural & fisheri \\
local & job & social & case & aye & deficit & passeng & school & avoid & war & peopl & art & servic & cull & fish \\
import & employ & prescript & amend & yemeni & spend & aircraft & children & treasuri & state & union & cultur & railway & cardiff & farm \\
mani & credit & practition & iss & noe & econom & expans & constit & hmrc & secretari & support & game & airport & england & insur \\
industri & support & transplant & way & plymyard & countri & island & mani & bonus & hous & import & beer & local & vaccin & agricultur \\
time & univers & mental & secretari & beg & rate & nois & support & regul & mani & british & club & passeng & languag & fishermen \\
invest & wage & pharmacist & public & abil & public & airlin & nhs & taxpay & militari & right & olymp & train & constit & rural \\
way & time & care & debat & abl & cut & gatwick & educ & incom & scotland & nation & shop & hs & tb & mesothelioma \\
hous & pay & asthma & mani & abolit & labour & termin & young & account & armi & uk & museum & network & cymr & dairi \\
iss & incom & patient & new & abrog & debt & baa & council & corpor & scottish & secretari & peopl & coast & assembl & welfar \\
countri & get & doctor & polic & absolut & job & ship & author & credit & support & world & alcohol & station & swansea & asbesto \\

\hline
\hline

\end{tabular}
\end{adjustbox}
   \caption{Top 15 words for each topic. Top panel displays first session, bottom panel displays last session.}
\end{table}
\end{landscape}




\newpage
\newpage

\clearpage

\subsection{Mapping The Speaker Activity}



In this section we  focus on how to distinguish and effectively visualize the activity of the different speakers, in other words how to map each speaker's role in the complex system that the UK parliament constitutes. The first piece of information about each speaker's relative importance is given by the \emph{length} of the speeches they give to the Parliament. We quantify such information in terms of the total number of words $w_i$ spoken by individual $i=1,\ldots,N$. As we will show in the following, the distribution of the speakers' verbosity appears to be strongly heterogeneous. 

Needless to say, not only the length of each speech, but the \emph{content} matters in determining the speaker's importance. In the previous section, DTM allowed us to identify the most significant $T$ topics debated in the UK parliament, and the contribution of each speaker to the different subjects over time. Hence, for each session it is possible to describe the activity of each speaker $i$ in terms of an activity vector

\begin{equation}
 {\bm p_i} = (p_i^{[1]}, ... ,k_i^{[T]}),
\end{equation}

where $\alpha=1,\ldots,T$ indicates the different topics and $p_i^{[\alpha]}$ indicates the fraction of time that the speaker $i$ spends talking about topic $\alpha$. Such values can be inferred by looking at the words used by the different speakers in their speeches and matching them to the topics to which they are considered to be strictly related. $p_i^{[\alpha]}$ can also be interpreted as the probability that, if we listen to a random speech of deputy $i$, she will be tackling topic $\alpha$. From the analysis of such vectors, we can identify different activity patterns. Stemming from this, it is for instance possible to obtain the similarity between the activity patterns of two speakers $i$ and $j$ by computing the Spearman correlation coefficient $\rho(\bm p_i, \bm p_j)$ \cite{spearman1904proof} or by mean of theoretic information measures, such as the Normalized Mutual Information $NMI(\bm p_i, \bm p_j)$ \cite{cover2012elements}. Such activity features can also be used to cluster, classify and visualize the speakers, as discussed in other sections of this manuscript. In general, a low (high) value of $p_i^{[\alpha]}$ indicates a low (high) engagement of the speaker $i$ with topic $\alpha$. However, to correctly take into account the speaker's contribution to a given topic, factors as the global importance of a given topic should be taken into account. If a topic is in general not strongly debated into the Parliament, a limited number of speeches concerning it should be sufficient to identify a speaker as a significant contributor. The significance of a topic $\alpha$ for a speaker $i$ can be easily determined by computing the Revealed Comparative Advantage (RCA) \cite{balassa1965trade}:
\begin{equation}
RCA_i^{[\alpha]}=\frac{\frac{p_i^{[\alpha]}}{\sum_{\alpha}p_i^{[\alpha]}}}{\frac{\sum_i p_i^{[\alpha]}}{\sum_{\alpha, i} p_i^{[\alpha]}}}.
\end{equation}  
If $RCA_i^{[\alpha]}>1$, i.e. $\frac{p_i^{[\alpha]}}{\sum_{\alpha}p_i^{[\alpha]}}>\frac{\sum_i p_i^{[\alpha]}}{\sum_{\alpha, i} p_i^{[\alpha]}}$, the fraction of time devoted by $i$ to topic $\alpha$ is greater than the average time devoted to the same topic by all speakers, and as a consequence topic $\alpha$ is a significant topic for speaker $i$. We notice that, given $p_i^{[\alpha]}<p_i^{[\beta}$, it is in general possible that $\alpha$ is a significant topic for speaker $i$ while $\beta$ is not, in spite of being, since RCA correctly discounts the individual activity with the overall importance of a topic.

\begin{figure}
\begin{center}
\includegraphics[scale=0.23]{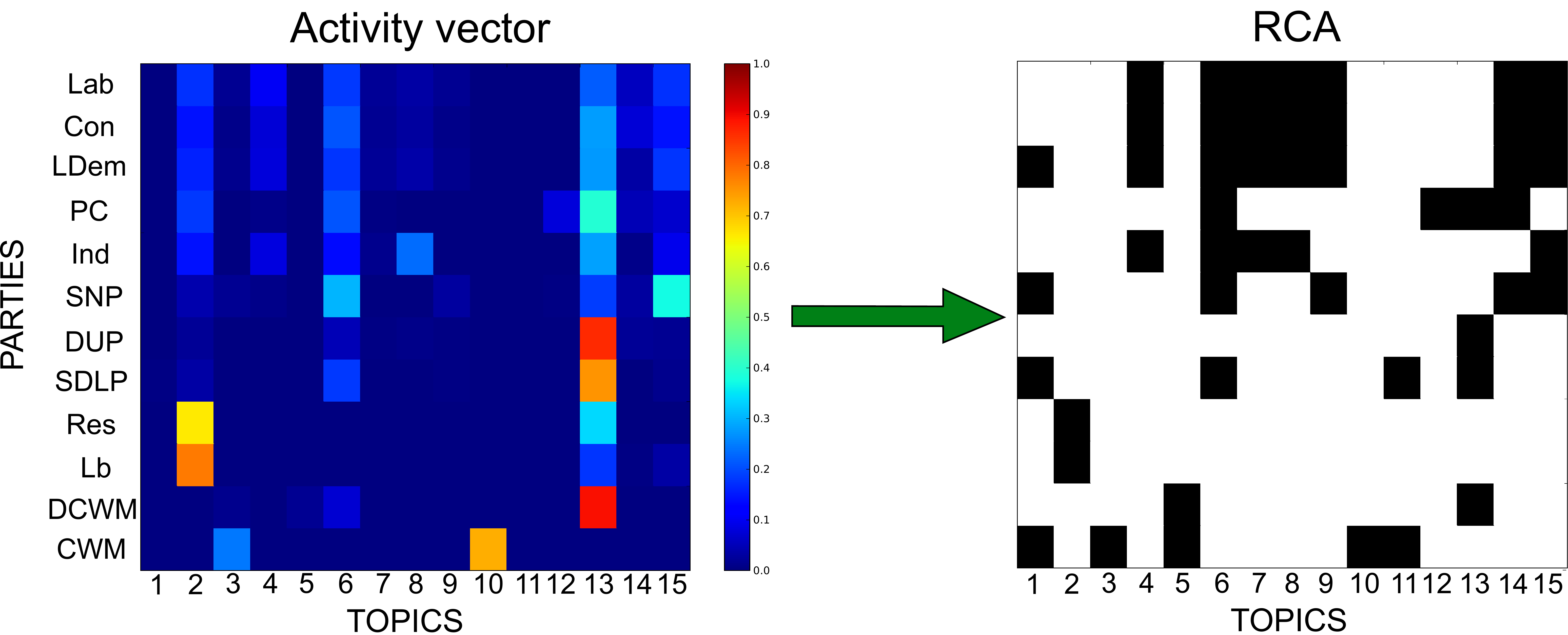}
\caption[]{In this figure we show the RCA indexes  across the different topics at the party level for session 5404 (black squares indicate $RCA=1$, white squares indicate $RCA=0$). On the left, a heatmap display the fraction of time that each party assigns to a given topic. To understand if a party is a significant contributor to a topic, however, it is necessary to compare such times with the average one assigned to the same topic by all parties. In such a way it is possible to unveil, for instance, how both Labour and Conservatives are significant contributor for topics 7,8,9, in spite of such topics not being among the most debated ones in the parliament.}
\label{fig:RCA}
\end{center}
\end{figure}

Starting from the individual activity vectors, the parliament activity vector ${\bm P_i}=(P_i^{[1]}, ... ,P_i^{[T]})$ can be easily obtained as the average of the speakers' activity weighted for their verbosity, i.e. $P_i^{[\alpha]}=\frac{\sum_i w_i p_i^{[\alpha]}}{\sum_i w_i}$. Analogously, activity vectors per parties can be obtained by performing the same operation and limiting the sum to the speakers belonging to a given group.
Consequently, it is also possible to obtain RCA indexes per party across the different topics. An example is shown in Fig. \ref{fig:RCA} for session 5404.

In general, a topic-by-topic exploration of the activity patterns of the speakers provides very detailed insights about the speakers' profile. As a drawback, given the large number $T$ of the considered topics, it is often difficult to visualize and evaluate it at a glance. An interesting information on the activity of each speaker is their capability to participate to the political debate in different topics. Such information can be synthetically evaluated by introducing the activity entropy $s_i$
\begin{equation}
s_i=-\sum_{\alpha}p_i^{[\alpha]}\ln p_i^{[\alpha]}.
\end{equation} 
By definition $s_i\ge 0$, with $s_i = 0$ only when the activity of a speaker is completely focused on a single topic, i.e. $p_i^{[\alpha]}=0$ $\forall \alpha=1,\ldots,N$ but one.
Greater values of $s_i$ indicates engagement in a variety of topics and are typical of generalist speakers which are able to deal with different political subjects. Conversely, low values point out to specialists, individuals who were able to construct their political careers thanks to their knowledge of specific areas, specialised skills and thematic persistence in their political speeches.

\begin{figure}
\begin{center}
\includegraphics[scale=0.4]{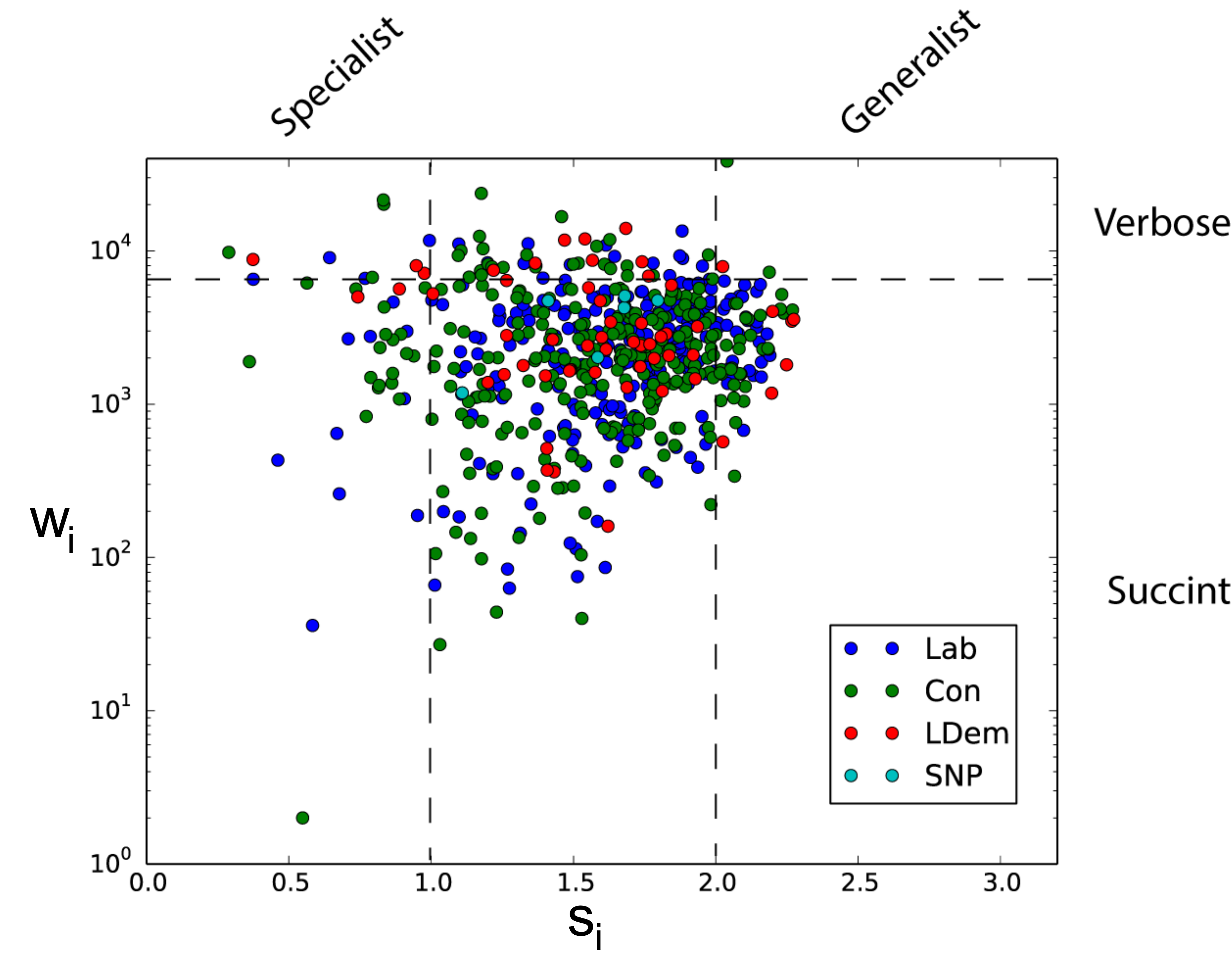}
\caption[]{In such figure we show the scatter plot of the verbosity $w_i$ against the activity entropy $s_i$ for the different speakers for four different parties. For all of them, the two variables do not appear to be correlated, indicating that the information they provide is complementary. Indeed, for a fixed level of $w_i$ speakers are found with very heterogeneous values of $s_i$ and viceversa. According to the different values of $(s_i,w_i)$ it is possible to characterize the speakers' activity according to six different regions, taking into account if they are succinct or verbose and if they are specialized or generalist in the topics they tackle.}
\label{fig:entropy}
\end{center}
\end{figure}

We are now ready to map the speakers' activity in the UK parliament by assigning each speaker their coordinates $(s_i,w_i)$ and representing them as dots in the plane Entropy-Verbosity. Results for the speakers in session 5404 are shown in Fig. \ref{fig:entropy} As shown, the two variables appear to be not correlated and provide two orthogonal insights towards the activity of the different individuals. Indeed, for a fixed level of $w_i$ speakers are found with very heterogenous values of $s_i$ and viceversa. For convenience, we divide the speakers in different categories according to their coordinates. In particular we have
\begin{itemize}
\item \emph{specialized} speakers for $s_i<1$;
\item \emph{mixed} speakers for $1\le s_i< 2$;
\item \emph{generalist} speakers for $s_i\ge 2$;
\end{itemize}
At the same time we differentiate between 
\begin{itemize}
\item \emph{verbose} speakers for $w_i<8*10^3$;
\item \emph{succint} speakers for $w_i\ge 8*10^3$;
\end{itemize}
Altogether, we are able to distinguish six different regions of political activity. For future developments, it would be interesting to evaluate the evolution of the activity over time for the different speakers and associate their position in such map with their electoral results and political membership.

\clearpage

\subsection{Topological Data Analysis}
\begin{figure}
\begin{center}
\includegraphics[scale=0.5]{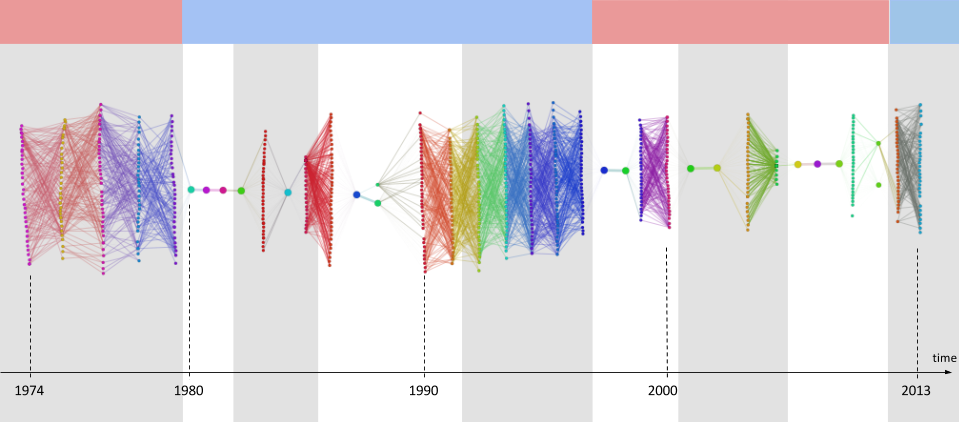}
\caption[]{In this graph each node is a different subset of politicians, and edges connect nodes that have politicians in common. The color of each node represents a different session. Sessions are identified by the year of their beginning (eg. session 4701 corresponding to year 1974-1975 is identified by 1974). The vertical bands distinguish between different parliament terms, the horizontal bands on the top indicate which party ruled during those years (Labour (Red), Conservative (Blue)). The number of clusters in the network varies significantly over time. For example it can be seen that fewer clusters where detected during periods of political stability during which Margaret Thatcher (1979-1990), and Tony Blair (1997-2007) held office.}
\label{fig:cluster_hoc}
\end{center}
\end{figure}

This section presents the results of the TDA applied to the results from the DTM analysis. Figure \ref{fig:cluster_hoc} presents the network constructed using the Mapper algorithm. Each node in the network represents a subset of politicians clustered according to Affinity Propagation algorithm. The network is colored so as to identify the different sessions to which each node belongs. An edge in the graph connects nodes that have politicians in common, which is why every node is only linked to nodes belonging to the previous or the subsequent session. 

The first thing one should notice is that the number of clusters in the network varies significantly over time. This is due to the clustering algorithm. 
To better study the results, we identified the political era which each session belongs to. Fewer clusters were detected during periods of political stability mainly in the years in which Margaret Thatcher (1979-1990), and Tony Blair (1997-2007) held office.

\begin{figure}
\begin{center}
\subfloat{\includegraphics[scale=0.35]{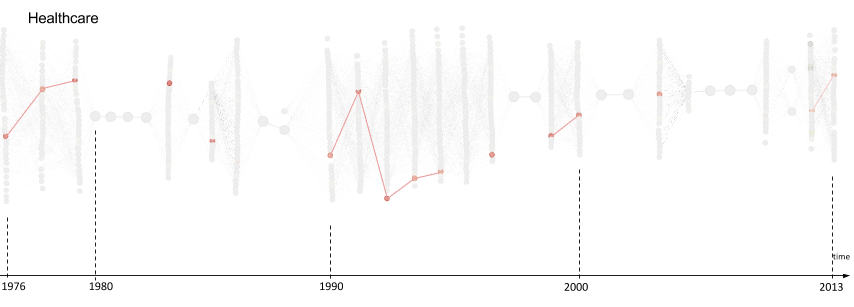}}
\subfloat{\includegraphics[scale=0.3]{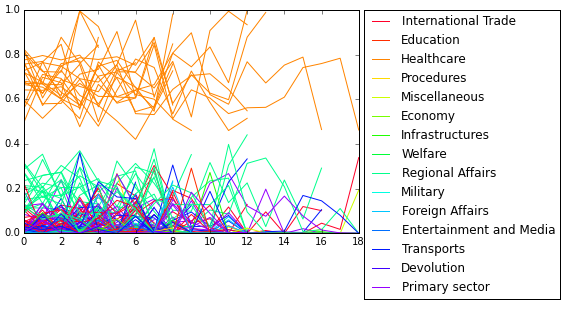}}
\caption[]{In this figure, only nodes with a median value for the topic `Healthcare' greater or equal than $0.70$ are selected. Nodes belonging to subsequent years are connected, which means that there are politicians that tend to specialize on the same topic over subsequent years. In the graph it is depicted the distribution of topics in each node selected in the network on the left. On the x-axis the politicians belonging to a node are represented, and on the y-axis the percentage of time talked on a certain topic over the session the node belongs to. It is clear from this graph, that the clusters are well defined and each of them contains politicians with a preference for talking about `Healthcare.'}
\label{fig:cluster_hocHealth}
\end{center}
\end{figure}

Another analysis we focused on was the evolution of similar clusters over time. As Figure \ref{fig:cluster_hocHealth} shows, clusters defined by a high frequency on a singular topic (Healthcare in the example showed in Figure \ref{fig:cluster_hocHealth}). Nodes belonging to subsequent years are connected, which means that there are politicians that tend to specialize on the same topic. In the example illustrated in Figure \ref{fig:cluster_hocHealth}, at least $10\%$ of politicians in connected nodes don't change their behavior over time, with picks of $64\%$ in session 5303 (2005). Being more or less inclined to change ones behavior does not seem to correlate with time. In future work we hope to verify if this variation in behavior might be influenced by external historical factors.

\newpage 

\newpage
\newpage
\newpage
\clearpage

\section{Concluding remarks and future work}

This paper has been an exploratory effort with the intention to pave the way towards a data driven understanding of politics and political institutions. We have proposed the use of dynamic topic modeling and topological analytical techniques to capture policy issues and their dynamics through unstructured political texts. We applied the methods on data referring to the UK House of Commons, uncovering two main results. First, we demonstrated the possibility of classifying Members of the Parliament according to their speeches' semantic content and verbosity. Second, we identified global patterns of political activity that suggest period of relative political cohesion or homogeneity. The findings generate a number of captivating research questions and hypotheses, such as: 

\begin{itemize}
	\item Can we systemically identify policy leaders and their effects on overall policy discussion and implementation process?
	\item Can we construct a robust framework that measures political structural stability by observing unstructured data?
\end{itemize}

In addition, our work lends itself to potentially useful policy applications. Results suggest that it feasible to track the performance and progress of individual MPs and parties with respect to specific policy issues, thus building the base for a data science framework to check the transparency and accountability of political agents. Other applications include the evaluation of novel indices of political stability and cohesion. \\

In future work we plan to extend our study by incorporating sentiment analysis. While a political agent's stance on a particular bill becomes a matter of public record -- crystallized in the stark binary `yea' or `nae' vote, data driven methods to profile their evolution to that position and their general disposition toward an entire policy topic are currently lacking \cite{Walker:2012:YEC:2364634.2364902}. Despite this, to further our aim of empirically investigating policy debate dynamics, we must account for some form of political opinion rather than relying solely on a speaker's topical content. Sentiment analysis of the traditional flavor will provide some information on political stance. We suspect that sentiment analysis on a per topic basis may even discriminate between parties taking opposite views on an topic -- although preliminary work using the Stanford Sentiment Treebank \cite{socher2013recursive} shows that overall sentiment in the House of Commons has a strong negative skew. We will analyze the sentiment results on a per-speaker, per-topic basis using the TDA method defined above, allowing us to answer questions regarding both individual and party level dynamics through the space of political positions. 

We also intend to investigate whether the topical and sentiment information extracted from unstructured data is a predictor of political success. More specifically, we will use Random Forests to study whether this kind of information predicts re-election of an individual, and perform an analysis of relative importance of these factors. 
Lastly, we will repeat topological data analysis analysis taking into consideration party membership of each politician, in order to track the evolution of the inner structure of the parties over time.

\section{Acknowledgments}
The authors acknowledge support from the Spanish MICINN Project CSD2010-00034 (SimulPast CONSOLIDER-INGENIO 2010).

\newpage
\begin{appendix} 

\section{Appendix: Model Selection: How to decide on the number of topics?}
 
Model selection, i.e. deciding the best values for the parameters of a model, is the key step to ensure the optimal performance of a model in balancing the two most important sources of error, bias and variance, resulting from the tradeoff between generalization capacity and flexibility of the model \cite{hastie2005elements}. One of the most common and powerful approaches to deal with this is cross-validation. When the data set does not contain a large number of instances, k-fold cross-validation is preferable in order to reduce the variability of the results. This approach involves creating a training and a validation set for each round, randomly dividing the set of observations into k groups approximately equal sized, called folds, and use one for validation and the remain k-1 for training \cite{breiman2001random}.
 
when using dynamic topic modeling, one of the important model parameters that needs to be fixed beforehand is the number of topics. To assess the performance of each trained model during the k-fold cross-validation scheme, a metric of goodness of the fit must be selected. In regression problems, MSE is the preferable metric. For DTM, there is no consensus or approach in the literature. Selecting this parameter value with a k-cross validation scheme is extremely computational expensive, that the usual way to approach this is to select the number of topics based on experience.
 
We have approached the problem differently. We have divided the observation data set in 37 time slices (each corresponding to a session) and selected 12 of them equally spaced in time. For each of the 12 slices, we have trained a LDA (Latent Dirichlet Allocation) model and then selected the optimal number of topics by the majority rule. We mimicked the essence of Random Forests \cite{hornik2011topicmodels}, with the difference that we gad LDA model instead of a Tree, and a training set selected ad-hoc instead of a bootstrapping sample. For the LDA, there are two âindustry-standardâ metrics generally used to assess the number of topics, Maximum Likelihood and perplexity. We used the log likelihood function to select the number of topics that yields the maximum likelihood of the model.

For each of the time slice considered, we did 5-fold cross-validation to assess the performance of 10,15,20,25 and 30 topics and chose the number of topics that maximized the log likelihood. We used the package "topicmodels for R \cite{elec}, and the functions LDA() with Gibbs sampling and logLik().
 
For 7 out of the 12 data sets, the log likelihood was maximized when the number of topics was 15, and the mean of the number of topics for the 12 data sets was 13.75; consequently we chose 15 as the optimal number of topics.

\end{appendix}

\newpage

\bibliographystyle{ieeetr}
\bibliography{polcomplex}

\begin{thebibliography}{10}

\bibitem{axelrod1981evolution}
R.~Axelrod and W.~D. Hamilton, ``The evolution of cooperation,'' {\em Science},
  vol.~211, no.~4489, pp.~1390--1396, 1981.

\bibitem{arthur1999complexity}
W.~B. Arthur, ``Complexity and the economy,'' {\em science}, vol.~284,
  no.~5411, pp.~107--109, 1999.

\bibitem{lampel2007role}
J.~Lampel and A.~Bhalla, ``The role of status seeking in online communities:
  Giving the gift of experience,'' {\em Journal of Computer-Mediated
  Communication}, vol.~12, no.~2, pp.~434--455, 2007.

\bibitem{battiston2015leveraging}
S.~Battiston, G.~Caldarelli, M.~D'Errico, and S.~Gurciullo, ``Leveraging the
  network: a stress-test framework based on debtrank,'' 2015.

\bibitem{cairney2012complexity}
P.~Cairney, ``Complexity theory in political science and public policy,'' {\em
  Political Studies Review}, vol.~10, no.~3, pp.~346--358, 2012.

\bibitem{easton1965framework}
D.~Easton, {\em A framework for political analysis}, vol.~25.
\newblock Prentice-Hall Englewood Cliffs, NJ, 1965.

\bibitem{hay1998structure}
C.~Hay and D.~Wincott, ``Structure, agency and historical institutionalism,''
  {\em Political studies}, vol.~46, no.~5, pp.~951--957, 1998.

\bibitem{young1981post}
R.~Young, ``Post-structuralism: An introduction,'' 1981.

\bibitem{sunkel1990neo}
O.~Sunkel and G.~Zuleta, ``Neo-structuralism versus neo-liberalism in the
  1990s,'' {\em Cepal Review}, 1990.

\bibitem{jervis1998system}
R.~Jervis, {\em System effects: Complexity in political and social life}.
\newblock Princeton University Press, 1998.

\bibitem{rhee2000complex}
Y.~P. Rhee, ``Complex systems approach to the study of politics,'' {\em Systems
  research and behavioral science}, vol.~17, no.~6, p.~487, 2000.

\bibitem{teisman2008complexity}
G.~R. Teisman and E.-H. Klijn, ``Complexity theory and public management: An
  introduction,'' {\em Public Management Review}, vol.~10, no.~3, pp.~287--297,
  2008.

\bibitem{geyer2010complexity}
R.~Geyer and S.~Rihani, {\em Complexity and public policy: a new approach to
  twenty-first century politics, policy and society}.
\newblock Routledge, 2010.

\bibitem{cairney2011understanding}
P.~Cairney, {\em Understanding public policy: Theories and issues}.
\newblock Palgrave Macmillan, 2011.

\bibitem{baumgartner2010agendas}
F.~R. Baumgartner and B.~D. Jones, {\em Agendas and instability in American
  politics}.
\newblock University of Chicago Press, 2010.

\bibitem{workman2009information}
S.~Workman, B.~D. Jones, and A.~E. Jochim, ``Information processing and policy
  dynamics,'' {\em Policy Studies Journal}, vol.~37, no.~1, pp.~75--92, 2009.

\bibitem{simon2000bounded}
H.~A. Simon, ``Bounded rationality in social science: Today and tomorrow,''
  {\em Mind \& Society}, vol.~1, no.~1, pp.~25--39, 2000.

\bibitem{henshel1987emergence}
R.~L. Henshel and W.~Johnston, ``The emergence of bandwagon effects: A
  theory,'' {\em The Sociological Quarterly}, vol.~28, no.~4, pp.~493--511,
  1987.

\bibitem{schmitt2008bandwagon}
R.~Schmitt-Beck, ``Bandwagon effect,'' {\em The International Encyclopedia of
  Political Communication}, 2008.

\bibitem{geyer2012can}
R.~Geyer, ``Can complexity move uk policy beyond ‘evidence-based policy
  making’and the ‘audit culture’? applying a ‘complexity cascade’to
  education and health policy,'' {\em Political Studies}, vol.~60, no.~1,
  pp.~20--43, 2012.

\bibitem{escher2011theyworkforyou}
T.~Escher, ``Theyworkforyou. com. analysis of users and usage for uk citizens
  online democracy,'' {\em UK Citizens Online Democracy}, 2011.

\bibitem{collins2009parallel}
C.~Collins, F.~B. Viegas, and M.~Wattenberg, ``Parallel tag clouds to explore
  and analyze faceted text corpora,'' in {\em Visual Analytics Science and
  Technology, 2009. VAST 2009. IEEE Symposium on}, pp.~91--98, IEEE, 2009.

\bibitem{ramage2009topic}
D.~Ramage, E.~Rosen, J.~Chuang, C.~D. Manning, and D.~A. McFarland, ``Topic
  modeling for the social sciences,'' in {\em NIPS 2009 Workshop on
  Applications for Topic Models: Text and Beyond}, vol.~5, 2009.

\bibitem{wallach2006topic}
H.~M. Wallach, ``Topic modeling: beyond bag-of-words,'' in {\em Proceedings of
  the 23rd international conference on Machine learning}, pp.~977--984, ACM,
  2006.

\bibitem{blei2012probabilistic}
D.~M. Blei, ``Probabilistic topic models,'' {\em Communications of the ACM},
  vol.~55, no.~4, pp.~77--84, 2012.

\bibitem{blei2006dynamic}
D.~M. Blei and J.~D. Lafferty, ``Dynamic topic models,'' in {\em Proceedings of
  the 23rd international conference on Machine learning}, pp.~113--120, ACM,
  2006.

\bibitem{hong2010empirical}
L.~Hong and B.~D. Davison, ``Empirical study of topic modeling in twitter,'' in
  {\em Proceedings of the First Workshop on Social Media Analytics},
  pp.~80--88, ACM, 2010.

\bibitem{gurciullo2014policy}
S.~Gurciullo, A.~Herzog, P.~John, and S.~Mikhaylov, ``Policy topics and their
  networks: Nlp and network analysis of uk house of commons debates,'' in {\em
  European Conference on Complex Systems Proceedings}, sep 2014.

\bibitem{carlsson_ams}
G.~Carlsson, ``Topology and data,'' {\em Bulletin of the American Mathematical
  Society}, vol.~46, no.~2, pp.~255--308, 2009.

\bibitem{singh}
G.~Singh, F.~M{\'e}moli, and G.~E. Carlsson, ``Topological methods for the
  analysis of high dimensional data sets and 3d object recognition.,'' in {\em
  SPBG}, pp.~91--100, Citeseer, 2007.

\bibitem{mapper}
P.~Lum, G.~Singh, A.~Lehman, T.~Ishkanov, M.~Vejdemo-Johansson, M.~Alagappan,
  J.~Carlsson, and G.~Carlsson, ``Extracting insights from the shape of complex
  data using topology,'' {\em Scientific reports}, vol.~3, 2013.

\bibitem{AP}
B.~J. Frey and D.~Dueck, ``Clustering by passing messages between data
  points,'' {\em science}, vol.~315, no.~5814, pp.~972--976, 2007.

\bibitem{spearman1904proof}
C.~Spearman, ``The proof and measurement of association between two things,''
  {\em The American journal of psychology}, vol.~15, no.~1, pp.~72--101, 1904.

\bibitem{cover2012elements}
T.~M. Cover and J.~A. Thomas, {\em Elements of information theory}.
\newblock John Wiley \& Sons, 2012.

\bibitem{balassa1965trade}
B.~Balassa, ``Trade liberalisation and “revealed” comparative advantage1,''
  {\em The Manchester School}, vol.~33, no.~2, pp.~99--123, 1965.

\bibitem{Walker:2012:YEC:2364634.2364902}
M.~A. Walker, P.~Anand, R.~Abbott, J.~E.~F. Tree, C.~Martell, and J.~King,
  ``That is your evidence?: Classifying stance in online political debate,''
  {\em Decis. Support Syst.}, vol.~53, pp.~719--729, Nov. 2012.

\bibitem{socher2013recursive}
R.~Socher, A.~Perelygin, J.~Y. Wu, J.~Chuang, C.~D. Manning, A.~Y. Ng, and
  C.~Potts, ``Recursive deep models for semantic compositionality over a
  sentiment treebank,'' in {\em Proceedings of the conference on empirical
  methods in natural language processing (EMNLP)}, vol.~1631, p.~1642,
  Citeseer, 2013.

\bibitem{hastie2005elements}
T.~Hastie, R.~Tibshirani, J.~Friedman, and J.~Franklin, ``The elements of
  statistical learning: data mining, inference and prediction,'' {\em The
  Mathematical Intelligencer}, vol.~27, no.~2, pp.~83--85, 2005.

\bibitem{breiman2001random}
L.~Breiman, ``Random forests,'' {\em Machine learning}, vol.~45, no.~1,
  pp.~5--32, 2001.

\bibitem{hornik2011topicmodels}
K.~Hornik and B.~Gr{\"u}n, ``topicmodels: An r package for fitting topic
  models,'' {\em Journal of Statistical Software}, vol.~40, no.~13, pp.~1--30,
  2011.

\bibitem{elec}
M.~Baxter, ``Electoral calculus: Historical data plots,'' 2015.

\end{thebibliography}

\end{document}